\documentclass[manuscript]{aastex}

\newcommand{\ind}[1]{_{\mathrm{#1}}}
\newcommand\deltanunu{\Delta\nu(\nu)}

\newcommand\dnup{\Delta\nu\ind{even}}
\newcommand\dnui{\Delta\nu\ind{odd}}

\shorttitle{KIC~11395018 and KIC~11234888}
\shortauthors{Mathur et al.}

\begin{document}


\title{Solar-like oscillations in KIC~11395018 and KIC~11234888 from 8 months of {\it Kepler} data}

\author{S. Mathur\altaffilmark{1}, R. Handberg\altaffilmark{2}, T.~L. Campante\altaffilmark{2,3}, R.~A. Garc\'ia\altaffilmark{4}, T. Appourchaux\altaffilmark{5}, T.~R. Bedding\altaffilmark{6}, B. Mosser\altaffilmark{7}, W. J. Chaplin\altaffilmark{8}, J. Ballot\altaffilmark{9}, O. Benomar\altaffilmark{5}, A. Bonanno\altaffilmark{10}, E. Corsaro\altaffilmark{10}, P. Gaulme\altaffilmark{5}, S. Hekker\altaffilmark{8,11},  C. R\'egulo\altaffilmark{12, 13}, D. Salabert\altaffilmark{12, 13}, G. Verner\altaffilmark{14,8},  T.~R. White\altaffilmark{6,15},  I.~M. Brand\~ao\altaffilmark{16}, O.~L. Creevey\altaffilmark{12,13}, G. Do\u{g}an\altaffilmark{2},   Y. Elsworth\altaffilmark{8}, D. Huber\altaffilmark{6},  S.~J. Hale\altaffilmark{8}, G. Houdek\altaffilmark{17}, C. Karoff\altaffilmark{2}, T.~S. Metcalfe\altaffilmark{1}, J. Molenda-\.Zakowicz\altaffilmark{18}, M.~J.~P.~F.~G.~Monteiro\altaffilmark{3}, M.~J. Thompson\altaffilmark{1}, J. Christensen-Dalsgaard\altaffilmark{2}, R.~L. Gilliland\altaffilmark{19}, S.~D. Kawaler\altaffilmark{20}, H. Kjeldsen\altaffilmark{2}, E.~V. Quintana\altaffilmark{21}, D.~T. Sanderfer\altaffilmark{22}, and S.~E. Seader\altaffilmark{21}}


\altaffiltext{1}{High Altitude Observatory, NCAR, P.O. Box 3000, Boulder, CO 80307, USA}
\altaffiltext{2}{Danish AsteroSeismology Centre, Department of Physics and Astronomy, University of Aarhus, 8000 Aarhus C, Denmark}
\altaffiltext{3}{Centro de Astrof\'isica and Faculdade de Ci\^encias, Universidade do Porto, Rua das Estrelas, 4150-762 Porto, Portugal}
\altaffiltext{4}{Laboratoire AIM, CEA/DSM -- CNRS - Universit\'e Paris Diderot -- IRFU/SAp, 91191 Gif-sur-Yvette Cedex, France}
\altaffiltext{5}{Institut d'Astrophysique Spatiale, UMR8617, Universit\'e Paris XI, Batiment 121, 91405 Orsay Cedex, France}
\altaffiltext{6}{Sydney Institute for Astronomy, School of Physics, University of Sydney, NSW 2006, Australia}
\altaffiltext{7}{LESIA, UMR8109, Universit\'e Pierre et Marie Curie, Universit\'e Denis Diderot, Obs. de Paris, 92195 Meudon Cedex, France}
\altaffiltext{8}{School of Physics and Astronomy, University of Birmingham, Edgbaston, Birmingham B15 2TT, UK}
\altaffiltext{9}{Institut de Recherche en Astrophysique et Plan\'etologie, Universit\'e de Toulouse, CNRS, 14 avenue E. Belin, 31400 Toulouse, France}
\altaffiltext{10}{INAF Osservatorio Astrofisico di Catania, Via S. Sofia 78, 95123, Catania, Italy}
\altaffiltext{11}{Astronomical Institute ``Anton Pannekoek'', University of Amsterdam, PO Box 94249, 1090 GE Amsterdam, The Netherlands}
\altaffiltext{12}{Universidad de La Laguna, Dpto de Astrof\'isica, 38206, Tenerife, Spain}
\altaffiltext{13}{Instituto de Astrof\'\i sica de Canarias, 38205, La Laguna, Tenerife, Spain}
\altaffiltext{14}{Astronomy Unit, Queen Mary University of London, Mile End Road, London E1 4NS, UK}
\altaffiltext{15}{Australian Astronomical Observatory, PO Box 296, Epping NSW 1710, Australia}
\altaffiltext{16}{Departamento de F\'isica e Astronomia, Faculdade de Ci\^encias da Universidade do Porto, Portugal}
\altaffiltext{17}{Institute of Astronomy, University of Vienna, A-1180, Vienna, Austria}
\altaffiltext{18}{Astronomical Institute, University of Wroc{\l}aw, ul. Kopernika 11, 51-622 Wroc{\l}aw, Poland}
\altaffiltext{19}{Space Telescope Science Institute, Baltimore, MD 21218, USA}
\altaffiltext{20}{Department of Physics and Astronomy, Iowa State University, Ames, IA 50011, USA}
\altaffiltext{21}{SETI Institute/NASA Ames Research Center, Moffett Field, CA 94035, USA}
\altaffiltext{22}{NASA Ames Research Center, Moffett Field, CA 94035, USA}



\begin{abstract}
We analyze the photometric short-cadence data obtained with the {\it Kepler Mission} during the first eight months of observations of two solar-type stars of spectral types G and F: KIC~11395018 and KIC 11234888 respectively, the latter having a lower signal-to-noise ratio compared to the former. We estimate global parameters of the acoustic (p) modes such as the average large and small frequency separations, the frequency of the maximum of the  p-mode envelope and the average linewidth of the acoustic modes. We were able to identify and to measure 22 p-mode frequencies for the first star and 16 for the second one even though the signal-to-noise ratios of these stars are rather low.
We also derive some information about the stellar rotation periods from the analyses of the low-frequency parts of the power spectral densities. A model-independent estimation of the mean density, mass and radius are obtained using the scaling laws. We emphasize the importance of continued observations for the stars with low 
signal-to-noise ratio for an improved characterization of the oscillation  modes. Our results offer a preview of what will be possible for many stars with the long data sets obtained during the remainder of the mission.

\end{abstract}





\section{Introduction}

Helioseismology has proved to be a powerful tool to directly probe the
interior of the Sun \citep[e.g.][]{2001NuPhS..91...73T,ThoJCD2003,2008SoPh..251...53C}. Thanks to years 
of continuous data, this tool provided a better understanding of the Sun and 
improved constraints on solar models 
\citep{JCDDap1996,GouKos1996,2009ApJ...699.1403B,2010Ap&SS.328...13S}. 
However, to establish the broader context and to make continued progress 
on stellar evolution theory, we need to study many stars across the HR diagram. Asteroseismology 
has progressed tremendously over the past decade, driven by 
ground-based observations and several satellite missions  \citep[see reviews by][]{1994ARA&A..32...37B,2001A&A...374L...5B,2002ASPC..259..464B,2008SoPh..251....3A,bedding08}. The CoRoT mission 
\citep{2006cosp...36.3749B} has provided data on a few stars showing 
solar-like oscillations 
\citep[e.g.][]{2009A&A...506...41G,2009A&A...506...33M,2010A&A...518A..53M}. 
For the stars observed with a high signal-to-noise ratio (SNR), the 
acoustic (p) modes could be unambiguously identified 
\citep{2009A&A...507L..13B,2009A&A...506...51B,2010A&A...515A..87D} and 
the signature of magnetic activity could be measured 
\citep{2010Sci...329.1032G}, complementing ground-based spectroscopic 
analyses \citep{1995ApJ...438..269B,2010ApJ...723L.213M}.

With the launch of the {\it Kepler Mission} 
\citep{2010Sci...327..977B,2010ApJ...713L..79K}, we now have continuous 
observations with a longer duration and higher precision, allowing the 
detection of more modes on many more stars. During the first year of {\it 
Kepler} observations, five stars were observed continuously for more than 
8 months. This is the first time that such long and continuous 
observations have been available. \cite{2010ApJ...713L.169C} used one month 
of data for several stars to demonstrate the asteroseismic potential of 
{\it Kepler}. We present an analysis of the 8-month-long time series for 
two solar-type stars, KIC~11395018 and 
KIC~11234888{\footnote {KIC~11395018 and 
KIC~11234888 are also known as Boogie and Tigger respectively within the Working Group \#1 (WG\#1) which is responsible for the analysis of solar-like stars}} while the analysis of two 
additional stars are presented in \cite{campante2010}.


According to the values of the stellar parameters ($T_{\rm eff}$, log$g$) given in the {\it Kepler Input Catalog}  \citep[KIC,][]{2010ApJ...713L.109B,2010ApJ...713L..79K}, these stars are expected to exhibit solar-like oscillations \citep{2011arXiv1103.0702C}. KIC~11395018 and KIC~11234888 have a {\it Kepler}-band magnitude of 10.8 and 11.9 respectively. Some spectroscopic analyses (Creevey et al., in preparation) show that KIC~11395018 is a G-type star with a $T_{\rm eff}$~=~5660~$\pm$~60~K (Pinsonneault \& An, in preparation). For KIC~11234888, the type has not been determined with certainty but it is likely to be a late F star, which has $T_{\rm eff}$~=~6240~$\pm$~60~K. Some features in their spectra suggest that they might be subgiants.

In the following 
Section we explain how the {\it Kepler} data have been processed for these 
two stars. In Section 3, we estimate the global parameters of the stars 
(background, rotation, mean large separation, mean small separation, mean linewidth). We characterize the p modes in Section 4, 
yielding lists of frequencies for both stars. Finally, we discuss and 
compare the results.

\section{Observations and Data Processing}


The data used in this work were collected by the {\it Kepler} photometer in the period from May 2009 to March 2010, corresponding to the first initial run (Q0), the first roll of one month long (Q1) and the next three rolls of three months each (Q2, Q3 and Q4). Unfortunately, due to the loss of  all the outputs in the third CCD-module on January 9, 2010, four CCDs were lost and in particular for these two stars, KIC~11395018 and KIC~11234888, we have only 20.87 days of measurements during the fourth roll. Thus, time series of 252.71 days ---with a short cadence of 58.85s \citep{2010ApJ...713L.160G}--- were available to the {\it Kepler} Asteroseismic Science Consortium  \citep[KASC;][]{2010arXiv1007.1816K} through the KASOC database\footnote{{\it Kepler} Asteroseismic Science Operations Center  \url{http://kasoc.phys.au.dk/}}
on these two targets. In the following, we refer to Q01234 as the full-length time series.

After the raw-pixel data were downlinked to the {\it Kepler} Science Office, light curves of the stars were created by calibrating pixels, estimating and removing sky background and extracting the time series from a photometric aperture \citep{vancleve2009}. 

Two types of light curves were available for each star: a {\it raw} one suffering from some instrumental perturbations, and a {\it corrected} one in which housekeeping data have been used to minimize those instrumental effects  during the Pre-search Data Conditioning (PDC) allowing the search for exoplanet transits \citep[e.g.][]{2010ApJ...713L..87J}. However, in some cases part of the low-frequency stellar signal (such as the one produced by starspots) can be filtered out. Therefore ---inside the WG\#1--- we have chosen to work with the {\it raw} data and developed our own method for the corrections \cite[see][]{2011arXiv1103.0382G}. 

\begin{figure}
\epsscale{.80}
\includegraphics[width=6.5cm, angle=90]{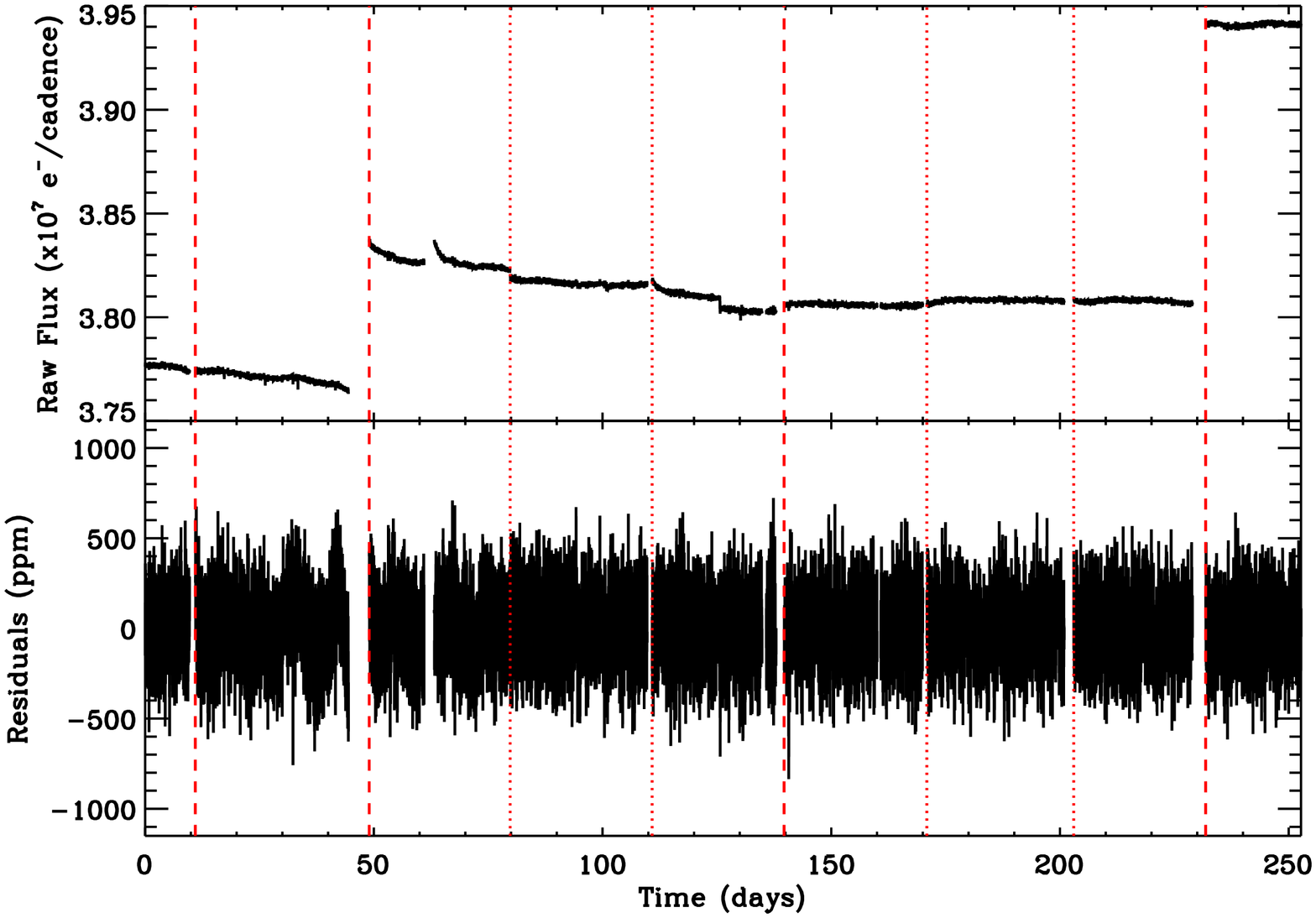}

\caption{Raw flux corrected for the outliers (top panel) and corrected data (bottom panel) for KIC~11395018. One point out of twenty has been plotted. The thick red dashed lines correspond to the beginning of each roll, while the thin red dotted lines correspond to a change of the sub-roll (e.g. Q2.2, Q2.3).\label{fig1}}
\end{figure}

The light curves were corrected for three types of effects: outliers, jumps, and drifts (top panel of Figure~\ref{fig1}). We have considered as outliers in the datasets the points showing a point-to-point deviation in the backward difference function of the light curve greater than $3\sigma$, where $\sigma$ is defined as the standard deviation of the backward difference of the time series. This correction removes $\sim$ 1$\%$ of the data points. Jumps are defined as sudden changes in the mean value of the light curve due, for example, to attitude adjustments or because of a sudden pixel sensitivity drop. Each jump has been validated manually. Finally, drifts are small low-frequency perturbations, which are in general due to temperature changes (after, for example, a long safe mode event) that last for a few days. These corrections are based on the software developed to correct the high-voltage perturbations in the GOLF/SoHO instrument \citep{GarSTC2005}. We fit a second or third order polynomial function to the region where a thermal drift has been observed after comparing several light curves from the same roll. Then, the fitted polynomial to the light curve is subtracted and we add another polynomial function (first or second order) ---used as a reference--- which has been computed from the observations done before and after the affected region. If the correction has to be applied on one border of the time series, only one side of the light curve is processed.

Once these corrections are applied, we build a single time series after equalizing the average counting-rate level between the rolls (bottom panel of  Figure~\ref{fig1}). A change of the average counting rate can happen inside a roll when a change of some instrumental parameters occurs. To do this equalization and to convert into parts per million (ppm) units,  we use a sixth order polynomial fit to each segment. 

As a consequence of these instrumental effects, the light curves from {\it Kepler} suffer from some discontinuities. For instance, they can be related to the pointing of the high-gain antenna towards the Earth --to send all the scientific data every month-- or to the rolling of the satellite that needs to maintain a proper illumination over the solar panels \citep{2010ApJ...713L..87J}. By taking into account these gaps, the {\it Kepler} duty cycle in the case of the observations of these two stars is limited to 93.45~\% of the time. On top of that, {\it Kepler} encountered some instrumental problems during these first eight months of measurements that we have corrected \citep{2011arXiv1103.0382G}. Therefore, the final duty cycle of the light curves are 91.36 and 91.34~\% for KIC~11395018 and KIC~11234888 respectively.






\section{Global parameters of the power spectrum density}

The power spectral density (PSD) of both stars is represented in Figure~\ref{fig2}. These have been obtained by applying the Lomb-Scargle algorithm \citep{1976Ap&SS..39..447L} to the ``corrected" data (see the previous section). In solar-like stars the spectrum at intermediate and low frequencies is dominated by the convective movements in their outer layers.

\subsection{Background parameters}


\begin{figure*}
\includegraphics[width=6.5cm, angle=90]{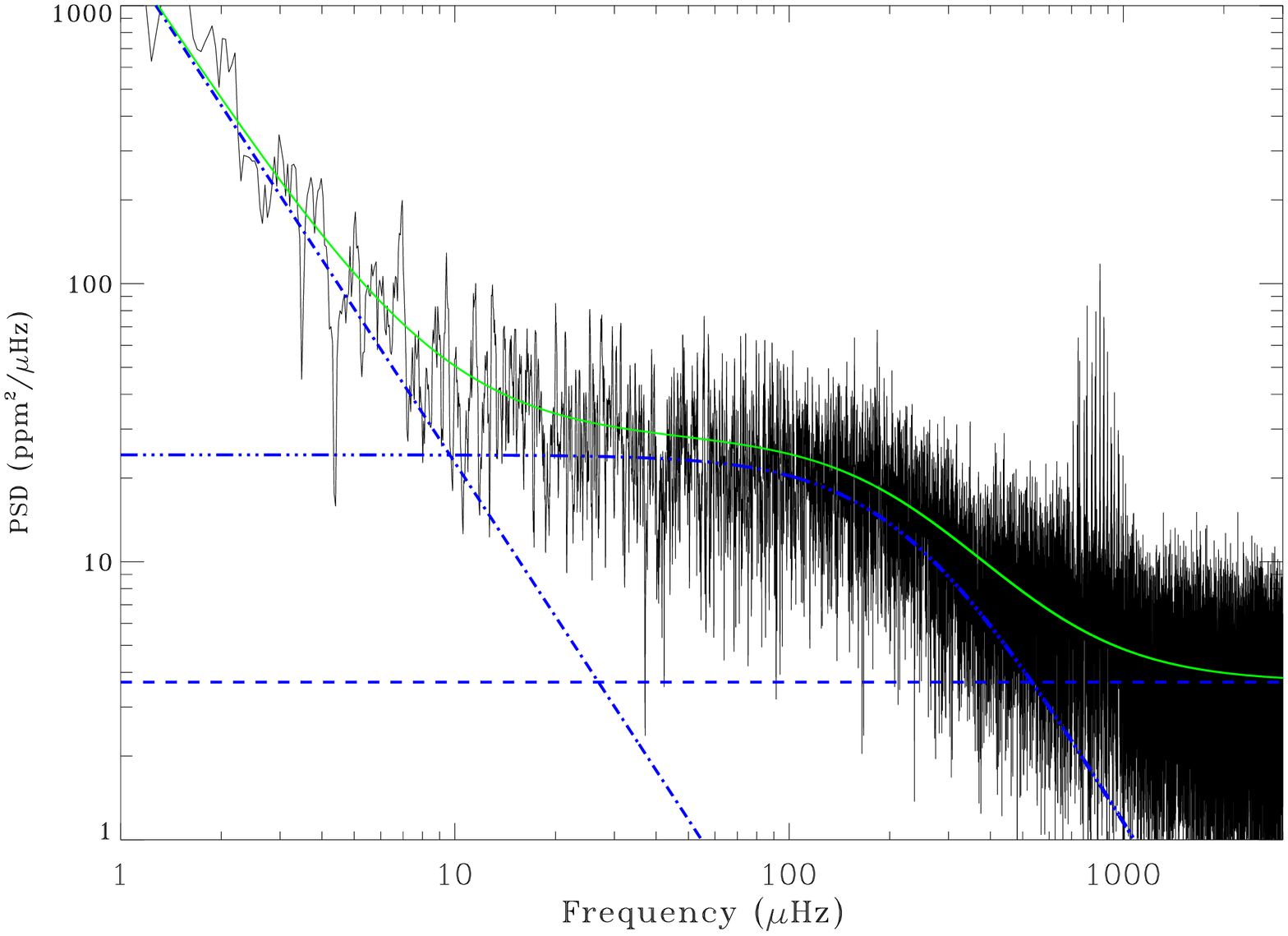}
\includegraphics[width=6.5cm, angle=90]{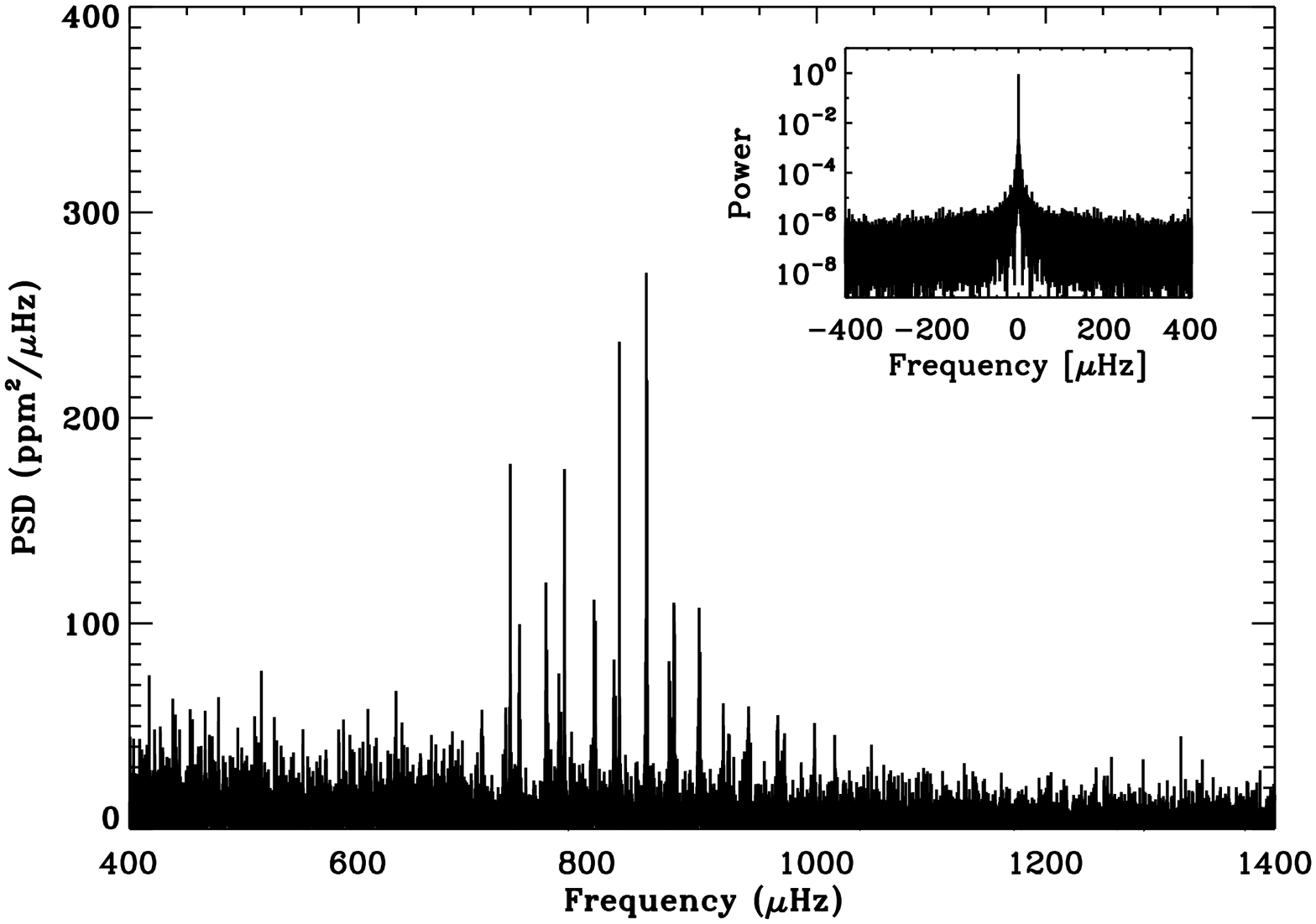} 

\includegraphics[width=6.5cm, angle=90]{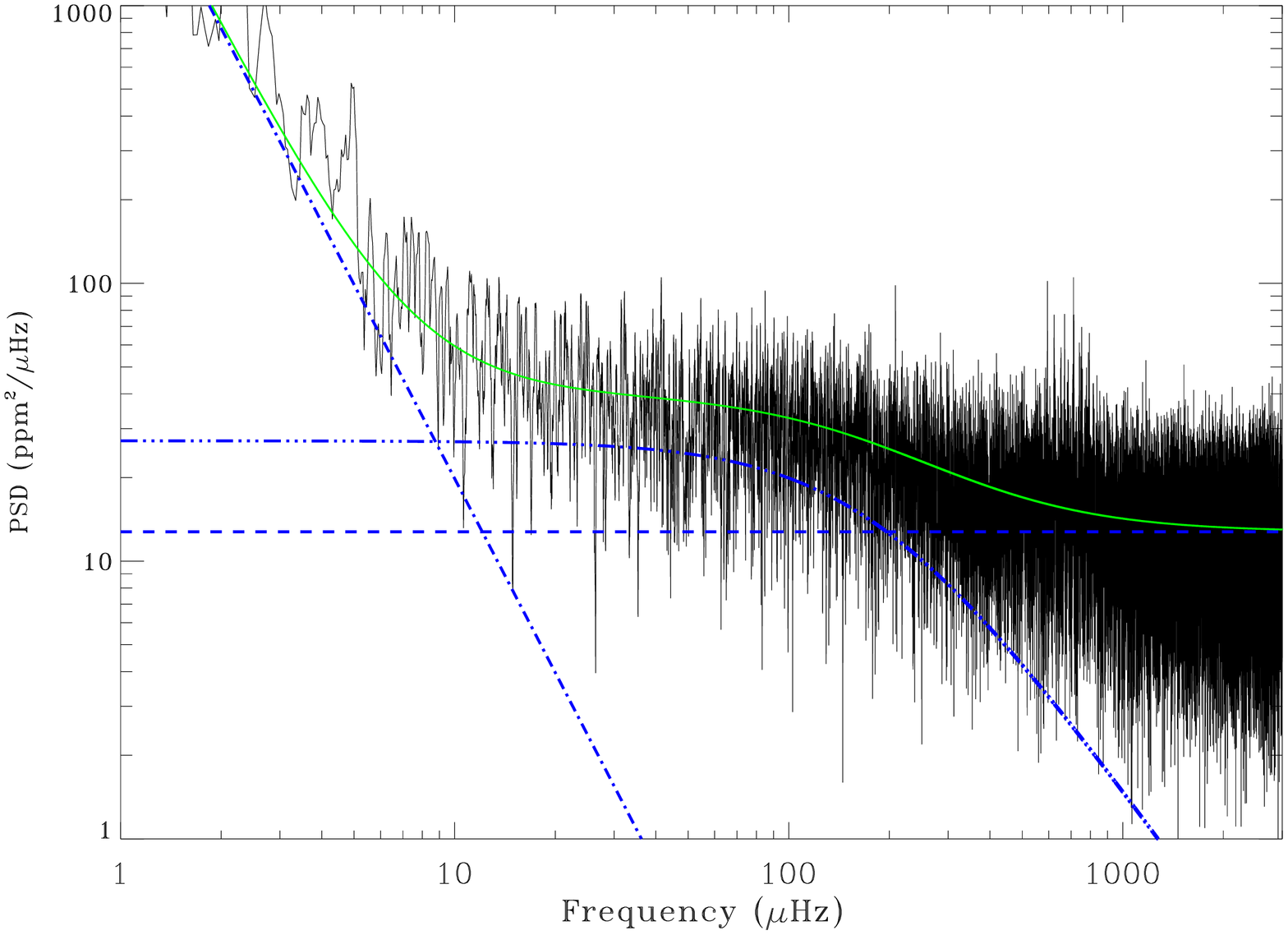}
\includegraphics[width=6.5cm, angle=90]{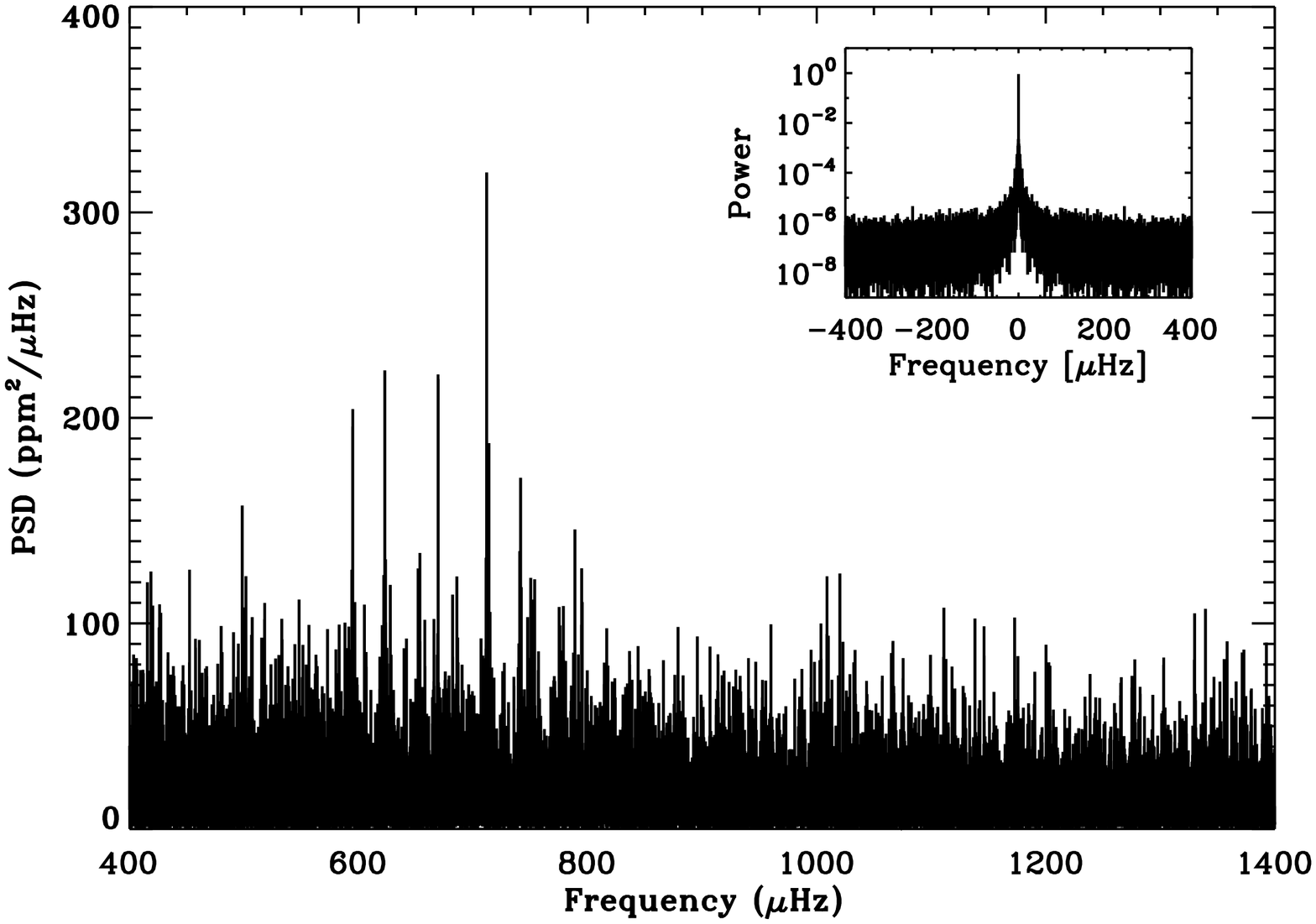} 

\caption{Left: PSD of KIC~11395018 (top) and KIC~11234888 (bottom) smoothed over a boxcar of 5 bins. The dashed line corresponds to the white noise level. The triple dotted-dashed line represents the Harvey law. The dotted-dashed line is the contribution of the power law. The solid line is the sum of all these contributions. Right: Zoom of the Power Spectrum Density of KIC~11395018 (top) and KIC~11234888 (bottom) using Q01234. The insets represent the power spectrum of the observation window. \label{fig2}}

\end{figure*}


The background of the star is modeled as:
\begin{equation}
B(\nu)=W+a\nu^{-b}+\sum_{i=1}^k 4 \frac{\tau_{i}\sigma_{i}^2}{1+(2\pi \nu \tau_{i})^{\alpha_{i}}},
\label{eq:bg}
\end{equation}
where $W$ is white noise, which models the photon shot noise, $a$ and $b$ are two parameters of a power law taking into account the effects of slow drifts or modulations due to the stellar activity, the instrument, etc., and $k$ is the number of Harvey laws that are used. $\sigma_i$, $\tau_i$ and $\alpha_i$ parameterize each of the $k$ Harvey-model contributions \citep{1985ESASP.235..199H}. 


In our case, equation~\ref{eq:bg} is fitted to the spectrum above 5~$\mu$Hz following the procedure described in \cite{2010A&A...511A..46M}. We initially assumed only one Harvey profile for the granulation. The measured characteristic time $\tau_c$, amplitude $\sigma_c$, exponent $\alpha_c$, and the photon  noise level for each of the two stars are listed in Table~\ref{tbl-0}.

We have verified that the results are not affected by the filtering of the time series. We performed the same fits to non-filtered spectra, where no polynomial function was subtracted. We recovered fully compatible results for the granulation background and the photon noise level, only the very low-frequency component is affected.

Finally, we have performed fits by adding an extra Harvey profile to model a faculae contribution \citep{2010AN....331..972K}. When one ensures that the faculae characteristic time $\tau_f$ is shorter than $\tau_c$ (which is also a free parameter), the fitting procedure converges to solutions where $\tau_f$ vanishes. We recover the previous values for the other parameters. Thus, we do not find any traces of faculae in the power spectrum of these two stars.

The background fits using only one Harvey law are represented in Figure~\ref{fig2} (left panel) for KIC~11395018 and KIC~11234888 (top and bottom panel respectively).





\subsection{Rotation period}

To determine the rotation period of the star, we investigate the low-frequency range of the PSD. Unfortunately, the standard procedure applied to process the data, as described in Section 2, filters out the lower part of the PSD below 1~$\mu$Hz. We have computed new merged time series where we applied a different high-pass filter by subtracting a triangular smoothed lightcurve with window sizes between 12 and 20 days. The influence of the instrumental drift in the very low frequency domain can thus be investigated.

\begin{figure}[h!]
\includegraphics[width=6cm, angle=90]{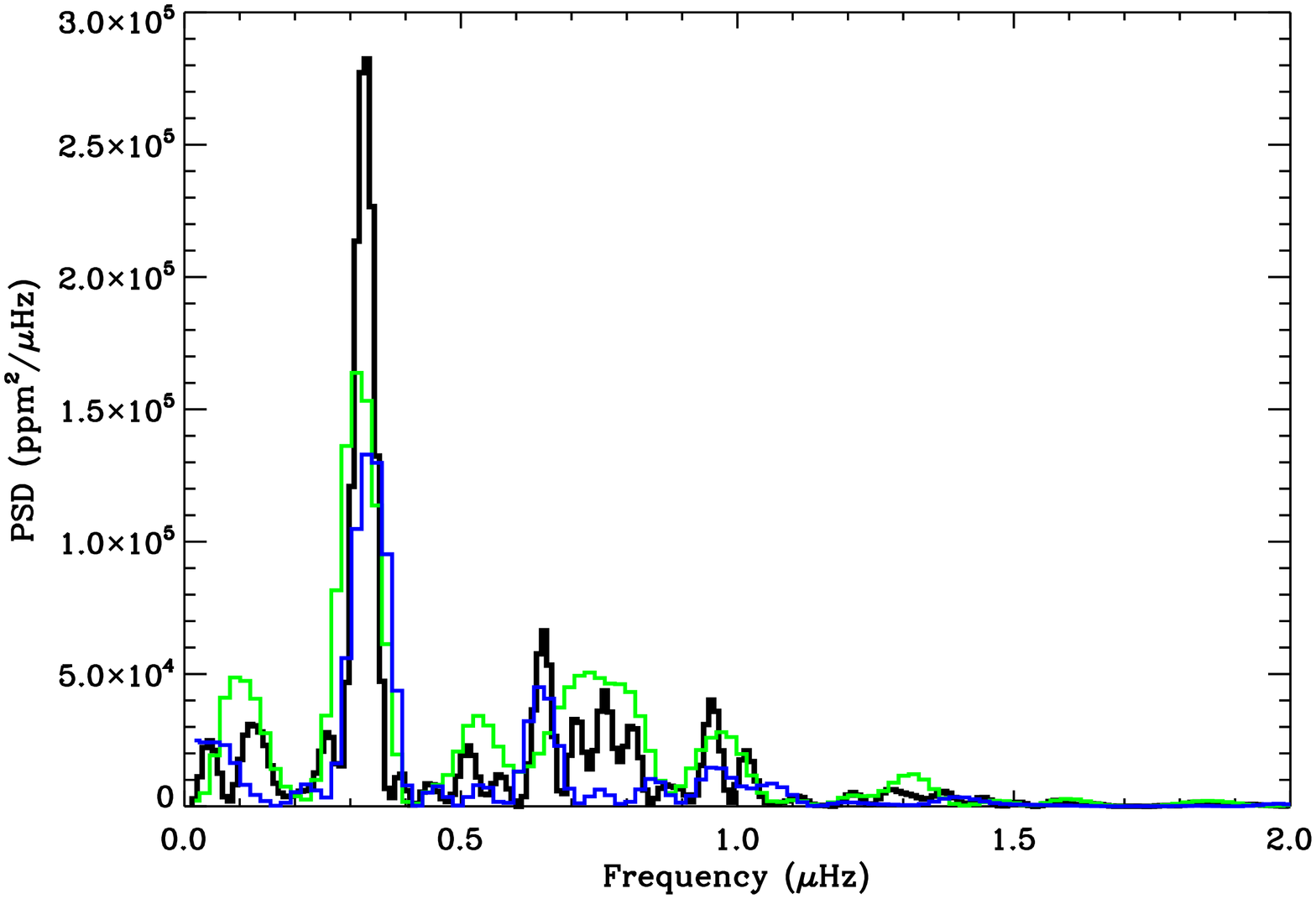}
\caption{\label{rot_boo} Zoom on the PSD at low frequency of KIC~11395018 oversampled by a factor of 5. In black the Fourier spectrum of the full-length series, In green, the PSD of the first half of the time series and in blue, the second half of the time series.}
\end{figure}

The data from Q0, Q1, and Q2.1 (the first month of Q2) show some instabilities due to instrumental effects. For KIC~11395018, we have computed the PSD of the full-length time series as well as of two subsets of 126 days each. A zoom of the low-frequency region of the PSD below 2~$\mu$Hz is shown in Figure~\ref{rot_boo}. In spite of the instabilities present in the first half of the time series, for the full-length and the half-length subseries the highest peak is at $\sim$~0.32~$\mu$Hz (Figure~\ref{rot_boo}), corresponding to a period of $\sim$36~days~$^{+6.04}_{-4.53}$~days, where we use the width of the resolution bin of the PSD of 0.0458~$\mu$Hz to compute the error bars. A closer inspection of the PSD of both subseries also implies that the main power is concentrated around this frequency but with a relatively more important contribution of the second harmonic $\sim 0.7$~$\mu$Hz. 

\begin{figure}[h!]
\includegraphics[width=6cm, angle=90]{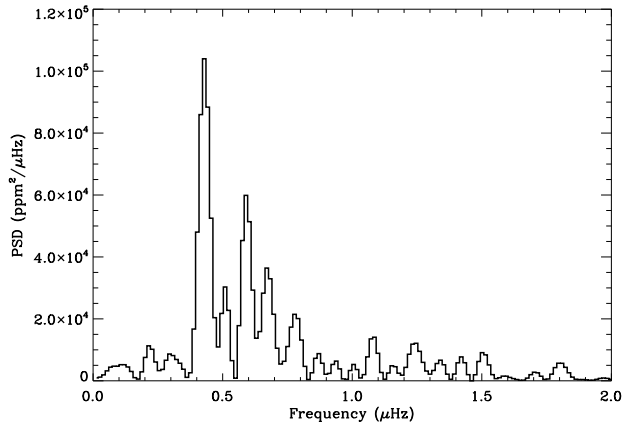}
\caption{\label{rot_tig} Zoom on the PSD at low frequency of KIC~11234888 from Q2.2 to Q4 data and oversampled by a factor of 5.}

\end{figure}

The case of KIC~11234888 is different as the Q0, Q1, and Q2.1 data are more unstable than for KIC~11395018 and the PSD of the first half of data is dominated by noise. Therefore, we have computed the PSD for time series from Q2.2 (the second month of Q2) to Q4, which is shown in Figure~\ref{rot_tig}. Here, we see one main peak at 0.42 and a group of peaks around 0.60~$\mu$Hz. This pattern would correspond to a rotation period of the stellar surface between 19~$^{+2.34}_{-1.88}$~days and 27~$^{+5.04}_{-3.69}$~days. The resolution is of 0.065~$\mu$Hz because of the shorter time-series analyzed for this star. The fact that we observe several high SNR peaks may suggest the existence of differential rotation on the stellar surface as observed in similar types of stars showing solar-like oscillations \citep{2009A&A...506..245M}. We notice that the highest peak is $\sim$2.5 times lower than the one in KIC~11395018. This could be due either to a low value of the inclination angle or to a less important surface activity leading to a smaller impact of the starspots in the light curves. However, for this star, we would need more data to confirm this rotation period.

Finally, we can say that these peaks detected at low frequency are likely to be of stellar origin as both stars were observed by the same CCD under the same conditions and show different periodicities.

\subsection{$p$-mode global parameters}

Several pipelines (AAU \citep{2010MNRAS.tmp.1125C}, A2Z \citep{2010A&A...511A..46M}, COR \citep{2009A&A...508..877M}, OCT \citep{2010MNRAS.402.2049H}, ORK \citep[tested in][]{2008ApJ...676.1248B}, QML \citep{2009A&A...506..435R}, SYD \citep{2009CoAst.160...74H}) analyzed the eight months of data to retrieve the global parameters: the mean large frequency separation ($\langle \Delta \nu \rangle$),  the position of the maximum amplitude ($\nu_{\rm max}$), the mean small frequency separation, $\langle \delta_{02} \rangle$, and the mean linewidth of the modes. 

$ \Delta \nu$ is the spacing between the frequencies of modes with the same degree ($l$) and consecutive radial orders ($n$) and depends directly on the sound travel time across the star. It is a very valuable parameter as it allows us to estimate the acoustic radius of the star and the mean stellar density \citep{1986ApJ...306L..37U}. Actually, the large separation is not a constant and varies with frequency, as shown in Section 3.3.2 so we also calculate the mean value of this quantity over a frequency range, $\langle \Delta \nu \rangle$.

The mean small separation, $\langle \delta_{02} \rangle$, is the mean value of the separation between two modes of consecutive radial orders and of degrees $l$ = 0 and 2. This quantity is sensitive to the structure of the stellar core and provides information about the age of the star.

\subsubsection{Estimation of the mean large separation and $\nu_{\rm max}$}
The different pipelines use different methods to compute $\langle \Delta \nu \rangle$ and $\nu_{\rm max}$: either the autocorrelation of the power spectrum or the autocorrelation of the time series. Having obtained comparable values for the two stars, we put one set of results in Table~\ref{tbl-1}.  For KIC~11395018, $\langle \Delta \nu \rangle$ is obtained in the range [570, 1140]~$\mu$Hz while for KIC~11234888, the mean value is calculated in the range [465, 935]~$\mu$Hz.

KIC~11234888 has a lower $\langle \Delta \nu \rangle$ and $\nu_{\rm max}$ than KIC~11395018.

The acquisition of longer time series allows better constraints on these global parameters with an improved precision. For KIC~11234888 the addition of Q34 (i.e. Q3 and Q4) to Q012 enabled us to make an estimate of $\langle \Delta \nu \rangle$ whereas with Q012 alone, we could not estimate it with certainty.
 
The $\langle \Delta \nu \rangle$ and $\nu_{\rm max}$ values obtained are consistent with the relationship derived by \citet{2009MNRAS.400L..80S}, viz. $\Delta \nu \propto \nu_{\rm max}^{0.77}$.



\subsubsection{Variation of $\Delta \nu$ with frequency}


As initially proposed by \citet{2009A&A...506..435R}, the
autocorrelation of the time series computed as the power spectrum of the power spectrum windowed with a narrow filter can
provide the variation of the large separation with
frequency, $\deltanunu$. The implementation of the method, the definition of the
envelope autocorrelation function (EACF) and its use as an
automated pipeline for the spectrum analysis have been addressed by
\cite{2009A&A...508..877M}. \cite{2010AN....331..944M} showed
that, with a dedicated comb filter for analyzing the power
spectrum, it is possible to obtain independently the values of the
large separation for the odd and even degree of the ridges (respectively $\dnui$
and $\dnup$). Then, proxies of the eigenfrequencies can be
derived by taking the highest peak in the region where the mode is expected from the variation of the large separation.

\begin{figure}[t]
\includegraphics[width=7.98cm]{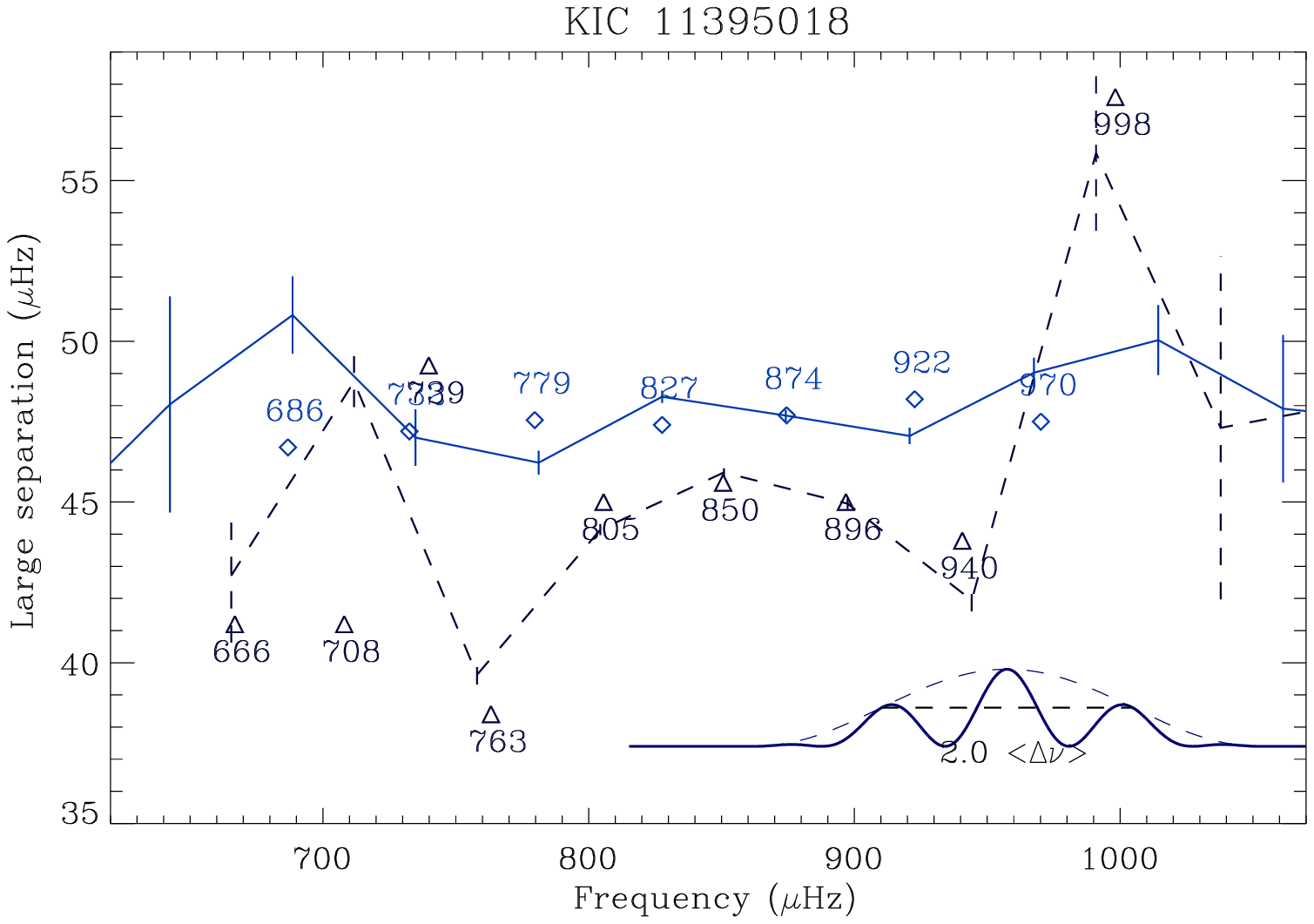}
\caption{$\dnup$ and $\dnui$ for KIC~11395018. The solid lines
(resp. dashed lines) correspond to the even (resp. odd)
ridge; 1-$\sigma$ error bars are given with the same linestyles.
We have superimposed the proxies of the eigenvalues derived from
the method: diamonds correspond to $l$ = 0, triangles to
$l$ = 1. The inset at the bottom right gives the size of the comb filter used for this star.
\label{fig_autodeltanuridge_11395018}}
\end{figure}

The values of $\dnup$ and $\dnui$ for KIC~11395018 are given in Figure~\ref{fig_autodeltanuridge_11395018}. The high SNR of the data for this star allows us to use a narrow filter for a detailed EACF analysis.The method shows the regularity of the even value  $\dnup$. It clearly emphasizes the lower values and the rapid variation of the large separation $\dnui$ compared to  $\dnup$. From this, we may expect an irregular \'echelle spectrum due to mixed modes \citep{2010AN....331..944M}.

\begin{figure}[t]
\includegraphics[width=7.98cm]{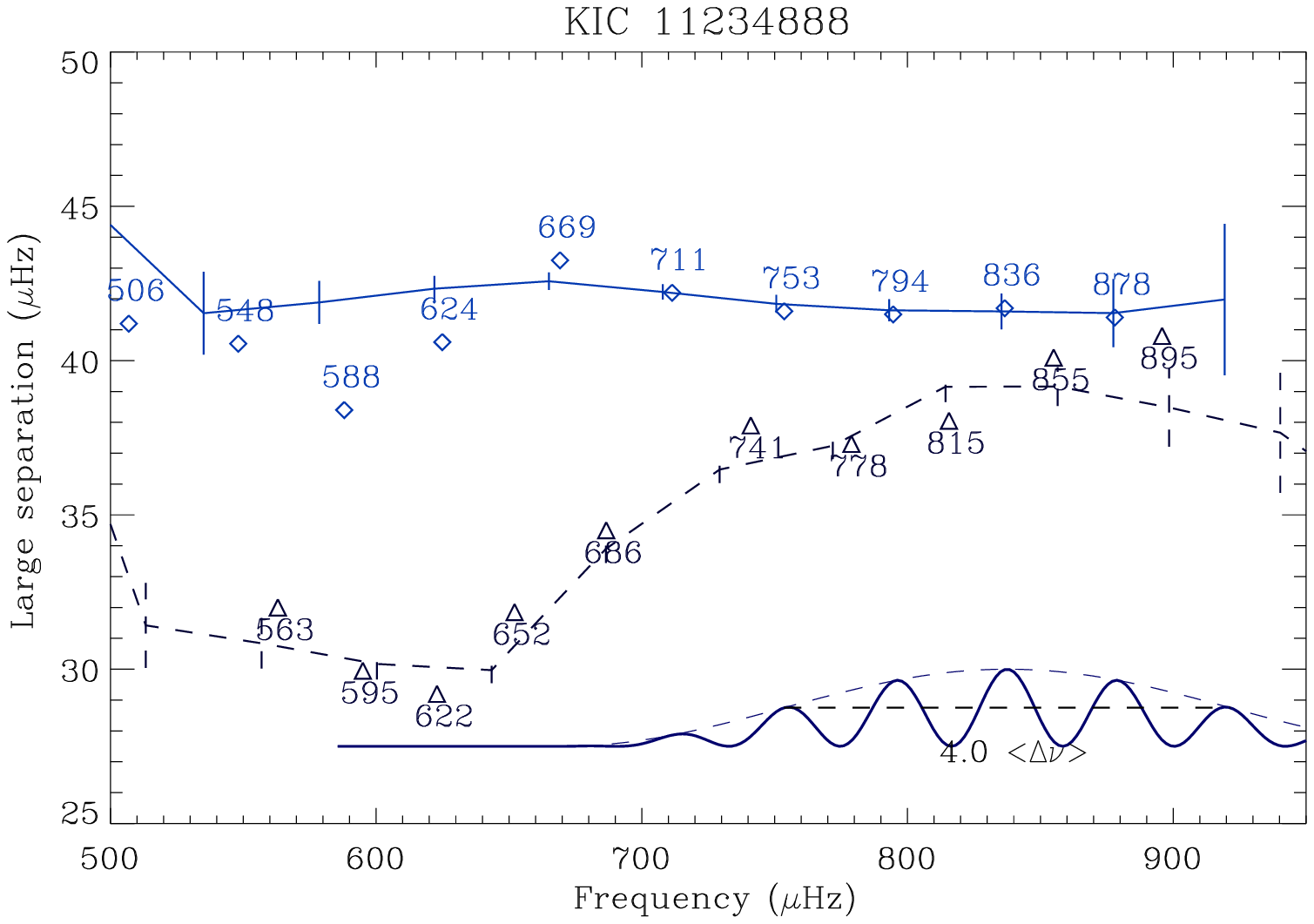}
\caption{Same as Figure~\ref{fig_autodeltanuridge_11395018} for KIC~11234888. \label{fig_autodeltanuridge_11234888}}
\end{figure}

Results for KIC~11234888 are given in Figure~\ref{fig_autodeltanuridge_11234888}. Compared to KIC 11395018, the lower SNR for this star requires the use of a broader comb-filter. However, the method shows unambiguously the differences between the even and odd values of the large separation and exhibits the low values of $\dnui$. Again, the presence of mixed modes is suspected.



\subsubsection{ Mean small separation and linewidth}


To obtain the mean small separation and mean linewidth, first the spectrum is smoothed over fifteen frequency bins. The spectrum is then folded using the value of the large spacing determined from fitting the modes so that the curvature of the \'echelle diagram is taken into account so that we could obtain vertical ridges.  For each degree $l$, we stack the peaks over seven large separations. This leads to an average or collapsed peak for each p-mode degree $l$. Two Lorentzian profiles were simultaneously fitted to the collapsed peaks of the $l$ = 0 and $l$ = 2  p modes.

The distance between the central position obtained for each Lorentzian with its error
is taken as the average small separation, $\langle \delta_{02} \rangle$,while the mean linewidth of the modes is computed from the fit to the average $l$~=~0 modes.  The non-linear least-square fitting procedure gives values for the linewidth and the central position of the collapsed peaks. Their associated error bars are calculated with the variance matrix. The 1-$\sigma$~error bar of the small separation is obtained by using a normal propagation of errors.

For KIC~11395018, we obtain a mean small separation of 4.12~$\pm$~0.03~$\mu$Hz in the range [650, 1000]~$\mu$Hz and for KIC~11234888, 2.38~$\pm$~0.19~$\mu$Hz in the range [550, 900]~$\mu$Hz. For the mean linewidth of the modes, $\langle \Gamma \rangle$, we obtain: 0.84~$\pm$~0.02~$\mu$Hz for KIC~11395018 and 0.86~$\pm$~0.06~$\mu$Hz for KIC~11234888.

\section{Characterization of the p modes}

\subsection{Fitting methods}

Eleven teams fitted the modes of these two stars.  We briefly
describe here the methods used by the different teams. Further details
may be found in Table~\ref{tbl-6}.

The methods used fall into three basic categories.  First, a majority
of teams adopted a maximum likelihood estimation approach \cite[MLE;
e.g.][]{AppGiz1998}, i.e., the best-fitting model of the
frequency-power spectrum was chosen by maximizing the likelihood of
that model. The model used was a sum of Lorentzian profiles describing
each oscillation mode, plus a background term parametrized as per the
descriptions in Section 3.1 \citep[see, e.g.][]{2008A&A...488..705A,
2009A&A...506...51B, 2009A&A...507L..13B, 2010ApJ...713..935B,2009A&A...506...41G,fletcher2010}. Because of the SNR of this particular set of data, it proved difficult to measure the asymmetry of the modes. Thus none of the teams include this parameter in the fit. However further efforts in the future may allow the measurement of the asymmetry in this and other {\it Kepler} stars. A few teams applied a derived version of this technique, the Maximum A Posteriori \citep[MAP;][]{2009A&A...506....7G}, where they add prior information to the MLE. The most common procedure of fitting is to do it globally
\citep{2008AN....329..485A}, where the entire frequency range is
fitted, meaning the entire set of free parameters needed to describe
the observed spectrum is optimized simultaneously. One fitter, however,
chose to fit the modes locally ($l$ = 0, 1, 2 (and 3) together) over one large separation, similar to what has been done for ``Sun-as-a-star'' helioseismology data
\citep{2004A&A...413.1135S}.


A second, smaller group of teams employed Bayesian Markov Chain Monte
Carlo algorithms \citep[MCMC; e.g.][]{2008CoAst.157...98B, 2011A&A...527A..56H}.  The MCMC
algorithm maps the probability density function of each free
parameter, and as such may be regarded as providing more robust
estimates of the confidence intervals on the parameters than is
possible from the standard MLE approach.

Finally, two teams applied a very different approach that did not
involve fitting a model to the observed frequency-power spectrum. They
adopted either a classical pre-whitening of the frequency-power
spectrum \citep[i.e. CLEAN algorithm, e.g.][]{2008ApJ...676.1248B}, or estimation of the
frequencies of the highest peaks in the smoothed spectrum.

Only five teams provided an estimate of the average rotational
splitting of the non-radial modes, and the angle of inclination of the
star (see Section~4.3.1 below).



\subsection{Methodology to select the frequencies}

A careful comparison of the different fitters' lists of frequencies yielded final
lists for each star, for use in future modeling work. The procedure
we adopted to compile the lists was a slightly modified version of
that described in \citet{2010ApJ...723.1583M}. 

At each \{$n,l$\} we compared the $N$ different estimated frequencies and identified and rejected outlying frequencies based on
application of the Peirce criterion \citep{1852AJ......2..161P,
1855AJ......4...81G}. 

The Peirce criterion is based on rigorous probability calculation and not on any ad-hoc assumption.
Hereafter we cite Peirce's explanation of his criterion: ``The proposed observations should be rejected when the probability 
of the system of errors obtained by retaining them is less than that of the system of
errors obtained by their rejection multiplied by the probability of making so many, and
no more, abnormal observations''   The logic calls for an iterative assessment of the rejection
when one or more datasets are rejected.  The iteration stops when no improvement is possible.

Following the work of \cite{1855AJ......4...81G}, we have implemented the Peirce criterion as follows:
\begin{enumerate}
	\item Compute mean $x_m$ and rms $\sigma$ deviation from the sample $x_i$
	\item Compute rejection factor $r$ from \cite{1855AJ......4...81G} assuming one doubtful observation
	\item Reject data if $|x_i-x_m| > r \sigma$
	\item If $n$ data are rejected then compute new rejection factor $r$ assuming $n+1$ doubtful observations
	\item Do step 3 to 4 until no data are rejected
\end{enumerate}

If the number of accepted frequencies was
greater than or equal to the integer value of $N/2$, we included that \{$n,l$\} on a
\emph{minimal} list for the star. This is the main difference from the previous procedure \citep{2010ApJ...723.1583M}, where all the teams had to agree on the mode to be put in the \emph{minimal list}. Inclusion on the \emph{maximal} list
demanded that there be at least two accepted frequencies for the mode
in question (i.e., we applied a more relaxed criterion for
acceptance). The additional modes in the maximal list are more uncertain and should be taken more cautiously but they can be used for instance to disentangle between two stellar models that fit the data. 

With the minimal and maximal lists compiled, the next step involved
computing the normalized root-mean-square deviation of the frequencies
of each of the $N$ teams, with respect to the average of the
frequencies of modes that appeared in the minimal list, i.e., we
computed
\begin{equation}
\sigma_{{\rm norm dev}, k} = \sqrt{\frac{1}{N_k}\sum_{n, l}\frac{
\left |
\nu_{n, l} ^k- \langle \nu_{n, l} \rangle
\right |^2}
{(\sigma_{n,l}^{k})^{ 2}}}
\end{equation}
where $k$ labels the team, $\nu_{n, l}^k$ and
$\sigma_{n,l}^k$ are the frequency and frequency uncertainty
returned by team $k$, $\langle \nu_{n, l} \rangle$ is the mean
value, over all teams, of the frequency of this mode, and $N_k$ is the
number of modes fitted by the team $k$ and that belong to the minimal list. The team with the
smallest $\sigma_{\rm norm dev, k}$ provided the frequencies which
populated the final minimal and maximal lists.

However, some issues remain with this more robust procedure. For instance, the team selected to provide the frequencies of the minimal and maximal lists might not have fitted one or several modes that should be in these lists. One way to overcome this issue would consist of asking the selected team to reanalyze the data using those additional modes. 
In the future, this procedure will be improved.


\subsection{Lists of frequencies}

\subsubsection{KIC~11395018}

Seven months of data have been analyzed by eleven fitters as described in Section 4.1.  Applying the selection methodology detailed in Section 4.2, we obtain a minimal list containing 22 modes, i.e. seven orders, as well as the fitter for whom we obtain the smallest normalized rms. 

Eight teams also analyzed the eight-month data sets. Unfortunately, these additional 21 days of Q4, obtained before CCD 3 was lost, did not allow us to identify any more modes in the lists. Table \ref{tbl-2} gives the minimal and maximal lists obtained for this star.  Outside the frequency range considered here, some more modes are still visible in the PSD and can therefore be fitted. The main issue encountered for these modes was to correctly identify their degree before doing the fit. In fact, the fits in these conditions become extremely dependent on the guess parameters and the identification given to the code beforehand. The existence of possible mixed modes complicates the task and we have decided to take a conservative position.  These frequency ranges will be explored in the future, when more data become available. 

\begin{figure}[h!]
\includegraphics[width=6cm, angle=90]{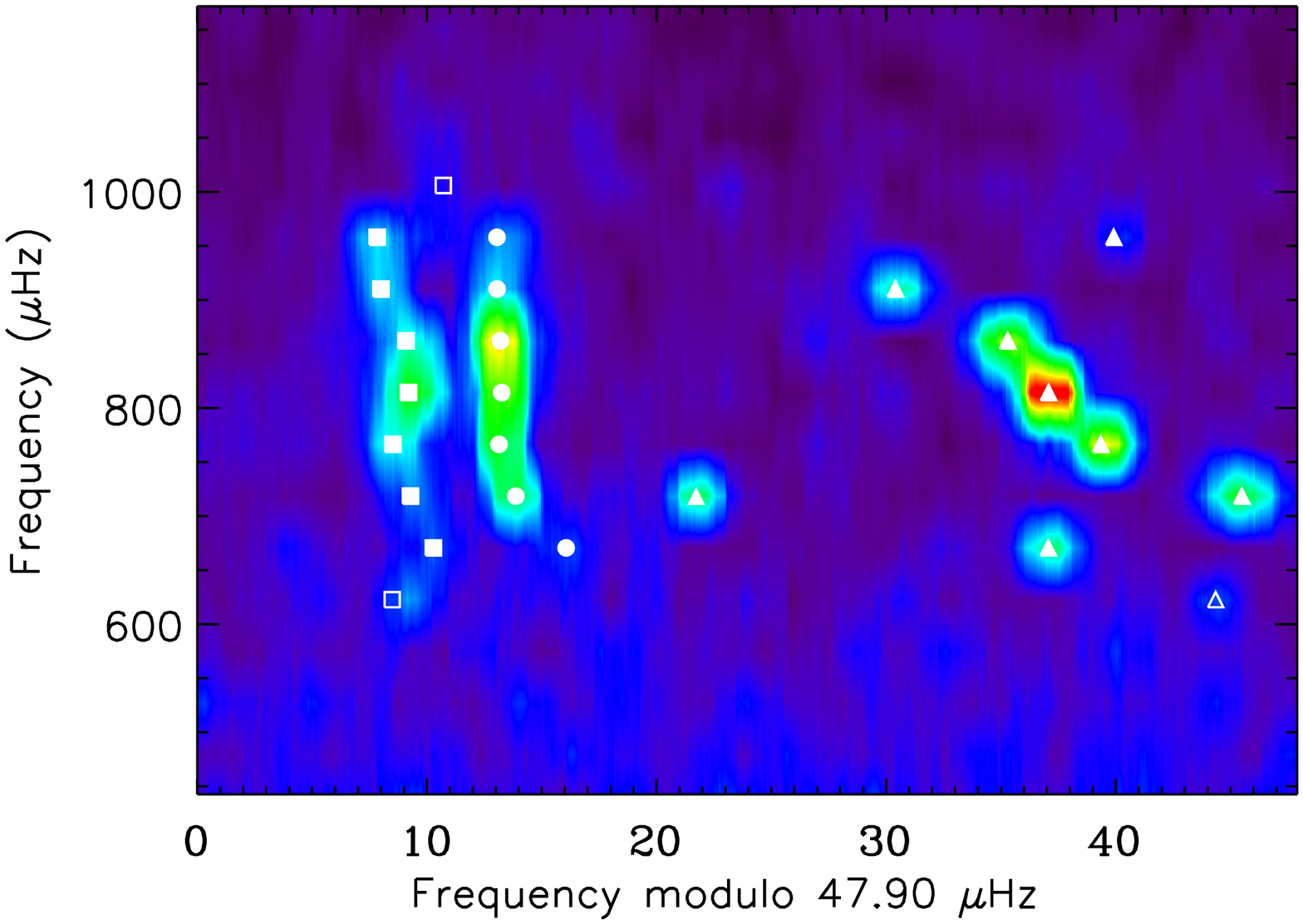}
\caption{\'Echelle diagram of the eight months of data with the minimal list of frequencies (filled symbols) and the maximal list (filled symbols plus open symbols) for KIC~11395018. The circles correspond to the $l$ = 0, the triangles to the $l$ = 1, and the squares to the $l$ = 2. \label{fig8}}
\end{figure}

Figure~\ref{fig8} represents the \'echelle diagram \citep{1983SoPh...82...55G} folded over 47.9~$\mu$Hz with the minimal and maximal lists, where frequencies have been obtained with the MAP method as a global fitting. We can clearly see the different ridges corresponding to the $l$ = 2, 0, and 1. While the ridges for the $l$ = 0 and 2 modes are almost straight, the dipole modes present a slope and two modes that have been shifted towards the right side of the \'echelle diagram near 763~$\mu$Hz and around 968~$\mu$Hz. This is a characteristic of the so called ``avoided crossing'' \citep{1977A&A....58...41A}, which is related to the presence of a mixed mode. This type of mode has the particularity of being supported both by pressure, like the acoustic modes and by gravity, like the gravity modes. They are thus sensitive to both the surface and the core of the star. The mixed mode is present simultaneously in p-mode and g-mode regions, hence the mixed character resulting in a shift of its frequency. 
These mixed modes are therefore very interesting as their characteristics are heavily dependent on how evolved the star is and thus they put additional strong constraints on the stellar  interior, providing a very good diagnosis on the age of the star \citep[][]{2009Ap&SS.tmp..241D,2010ApJ...723.1583M}. 

For the fitter with the lowest normalized rms, the mixed mode and the mode around 667~$\mu$Hz have been fitted separately, i.e. with the p modes of the previous large spacing, instead of globally. Indeed, if the modes are fitted globally, one assumes that the modes have a common linewidth and amplitude at every order (\{$l$ = 0, $n$\}; \{$l$ = 1, $n$\}; \{$l$ = 2, $n$-1\}). Thus the first $l$ = 1 and the mixed modes are treated as ``single modes". The fitting code fixes all the parameters to the values obtained for the modes previously fitted globally and then fits these two modes independently with all the parameters free.

The maximal list contains three more modes at higher and lower frequency.  One interesting peak is the one around 1016~$\mu$Hz. It is located between the two ridges $l$ = 0 and 2, and is identified with a high SNR in the PSD. However, its identification remains uncertain as it could be either an $l$ = 0 or an $l$ = 2. Moreover the presence of a mixed mode in this region is not excluded and could be responsible for this bumped peak. Note that all the modes of the minimal list were also detected by the EACF method of Section 3.3.2.

Finally, the fitter A2Z RG reported a problem while fitting the splittings and the inclination angle, which are  parameters that are highly correlated \citep[e.g.][]{2003ApJ...589.1009G,2006MNRAS.369.1281B}. Indeed, with the whole dataset, the fit converged to null values, while with the Q0123 data, the fit managed to converge to a value for the splitting of $\nu_\textup{s} =0.49~\pm~0.14~\mu$Hz and an inclination angle of $i$~=~36.83~$\pm$~10.78$^{\circ}$. So far, the reasons for obtaining null values remain unclear but it could either be related to a combination of the stochastic excitation, the width, and the splittings of the modes or due to a shift of the modes that prevents us from properly distinguishing the individual components of the modes.

\begin{figure}[htb]
\includegraphics[width=8.5cm]{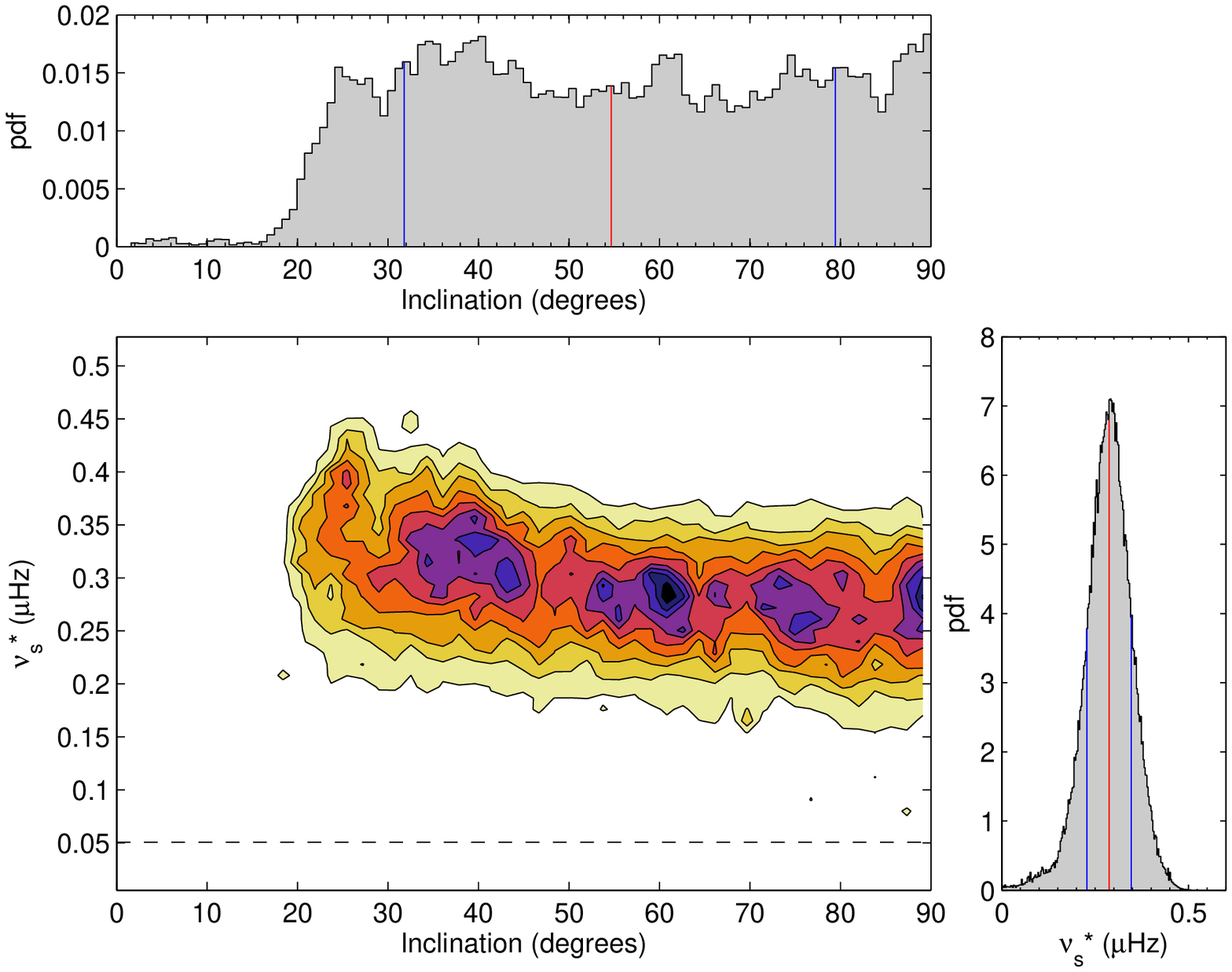}
\caption{Probability distribution of the splittings and the inclination angle for KIC~11395018 (bottom left panel). Dark colors refer to higher probability values compared to light colors.The projected rotation is given in the right panel while the inclination angle is represented in the top left panel. In each of these two panels, the red line corresponds to the median of the distribution and the blue ones indicate the 68\% confidence interval around it.
The dashed line in the contour-plot indicates the resolution in the power spectrum.\label{fig9}}
\end{figure}

On the other hand, the MCMC fitting method used by the fitter AAU gave an estimation of the projected rotational splitting $\nu_\textup{s}^* = \nu_\textup{s}\sin(i)$ of 0.29$\,\pm 0.06 \:\mu\textup{Hz}$ and an inclination angle $i \gtrsim 20^\circ$, with a confidence level of 68~$\%$ by analyzing Q0123 data. Indeed \cite{2006MNRAS.369.1281B} demonstrated that it is more robust to fit the projected splitting, $\nu_\textup{s}^*$, and the inclination angle instead of fitting the usual combination of splitting and inclination angle. The MCMC method fits the spectrum including rotational splittings as described by \cite{2003ApJ...589.1009G}, and using a set of priors. The projected rotational splittings, $\nu_\textup{s}^*$, and the inclination angle, $i$, are free parameters. Uniform priors (equal probability for all values) are used for both the inclination and $\nu_s$. The inclination is searched in the range [0-90] degrees while $\nu_s$ is assumed to be in the range [0-2]~$\mu$Hz. Figure~\ref{fig9} shows the PDF (probability density functions) of both fitted parameters. These preliminary results need to be confirmed with longer datasets. 


\subsubsection{KIC~11234888}

\begin{figure}[htb]
\includegraphics[width=6cm, angle=90]{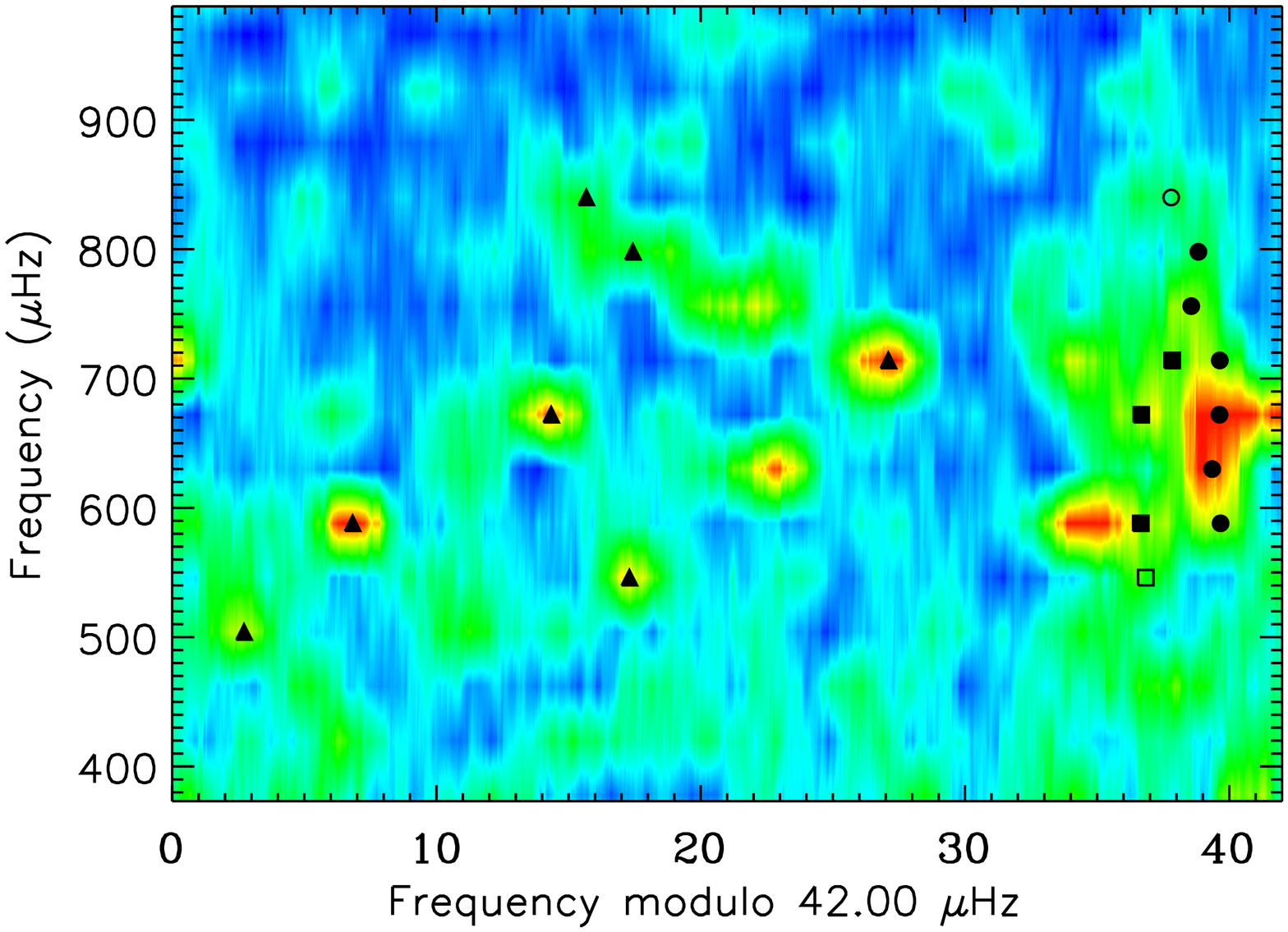}

\caption{\'Echelle diagram of the eight months of data with the minimal and maximal lists of frequencies for KIC~11234888. Same legend as Figure~\ref{fig8}.  \label{fig10}}
\end{figure}

KIC~11234888 has a more complex oscillation spectrum. Furthermore, it has a lower SNR, which results in a noisier \'echelle-diagram (Figure~\ref{fig10}). To compute the minimal and maximal lists of frequencies, we had to discard the results of two teams out of the six who fitted the modes, as one of them could not identify any modes and the other only fitted  a very small number of modes, leading to very short frequency lists. As a consequence, we have used the results of only four fitters to build the final minimal and maximal lists. Applying the method described in Section~4.2, we also selected the fitter with the lowest normalized rms. 

The frequencies obtained by looking for the highest peaks in the power spectrum are listed in Table~\ref{tbl-3}. The minimal list contains a total of 16 modes, while the maximal list has two more modes. Figure~\ref{fig10} illustrates the minimal and maximal lists overplotted on the \'echelle diagram. 

Like the previous star, KIC~11234888 was analyzed with the Q0123 (which corresponds to Q01234 data without the last quarter, Q4) and Q01234 data. By adding 21 days of data, two modes are discarded in the lists: an $l$ = 1 at 901~$\mu$Hz and an $l$ = 2 at 792~$\mu$Hz.

In Figure~\ref{fig10}, we can see the ridges for the $l$ = 0, 1, and 2 modes. But for the $l$ = 0 and 2, the SNR is quite low and the modes are not all obviously distinguishable.  For the ridge of the $l$ = 1 modes, we notice a very interesting structure showing several avoided crossings, due to the presence of several mixed modes. 

We can see that there are some regions in the \'echelle-diagram where we see some power but they are not listed as modes. The power present in the $l$~=~1 ridge was rejected according to the Peirce criterion. 

We also notice that the modes selected in the minimal list were also detected by the EACF method. 

Unfortunately for this star, no value for the splittings and inclination angle were provided.

We need more data and modeling to better understand this pattern and detect the $l$ = 0 and 2 modes with more certainty.




\section{Discussion and Conclusions}

In this work we analyzed the light curves corrected in a specific way of two solar-like stars, a G star, KIC~11395018, and an F star, KIC~11234888, observed by the {\it Kepler Mission} during $\sim$~8 months. Unfortunately, both stars were located on the same CCD that was on the module that broke in January 2010. Thus, only 8 months of continuous measurements were available for asteroseismic investigations. {\it Kepler} will continue observing these stars for at least 2.5 years. Because these stars will be regularly observed by the broken CCD, the data will contain periodic gaps of three months in length. We will thus have to deal with these gaps either by using subseries of 270 days, by averaging the power spectra of subseries, or by removing the fundamental peak and its harmonics in the PSD corresponding to the window function. 

Seven pipelines analyzed these data using different methods to retrieve the global parameters of the two stars. We obtained the mean large separation, $\langle \Delta \nu \rangle$ = 47.76~$\pm$~0.99~$\mu$Hz and $\langle \Delta \nu \rangle$ = 41.74~$\pm$~0.94~$\mu$Hz for KIC~11395018 and KIC~11234888 respectively. As the mean large separation is proportional to the average density of the star, we can deduce that KIC~11395018 is more dense.

KIC~11395018 is expected to be a post-main-sequence star due to the clear avoided crossings. Moreover, the presence of larger number of mixed modes in the \'echelle diagram of KIC~11234888 suggests that it is more evolved than KIC~11395018.

We determined the mean values of the small separation, $\langle \delta_{02} \rangle$ = 4.12~$\pm$~0.035~$\mu$Hz and 2.38~$\pm$~0.19~$\mu$Hz respectively for KIC~11395018 and KIC~11234888, as well as the mean linewidth of the modes, $\langle \Gamma \rangle$ = 0.84~$\pm$~0.02~$\mu$Hz and 0.86~$\pm$~0.06~$\mu$Hz respectively. Note that though these stars have different $T_{\rm eff}$, $\sim$5660~K for KIC~11395018 and $\sim$6240~K for KIC~11234888, we find similar values of linewidths, which is not expected according to \cite{2009A&A...500L..21C} who concluded that the mean linewidth of the modes scales like $T_{\rm eff}^{-4}$.

An analysis of the convective background for both stars demonstrated that KIC~11234888 exhibited the largest granulation time scale of the two stars. No signature of faculae was found in either of the stars.


Several teams (up to eleven) analyzed the time series of the two stars with different methods and provided p-mode frequencies. After applying a revised procedure to these different sets of frequencies, described in Section 4.2, we selected the modes on which the different teams agreed to create the {\it minimal} and {\it maximal} lists of frequencies for both stars. We identified 22 p modes in the range 600 to 1000~$\mu$Hz for KIC~11395018 and 16 p modes in the range 500 to 900~$\mu$Hz for KIC~11234888. These frequencies are in agreement with the modes detected by the EACF method.

The rotational splittings have been tentatively measured for KIC~11395018 by four teams. The MCMC gave projected splitting of $\sim$~0.29~$\mu$Hz. Combined with the measurement of the rotation period of 36 days obtained from an analysis of the low-frequency region of the PSD, we improve the inclination angle constraint to $i~\ge$~45$^{\circ}$. For KIC~11234888, which seems to show evidence of a differential rotation, we need more data to constrain the splittings and the inclination angle. 

Note that for both stars no $l$~=~3 modes have been detected. Indeed degrees higher than 2 are difficult to detect because of cancellation effects in spatially unresolved observations. However, clear evidence of $l$~=~3 modes has already been seen in red giants observed by {\it Kepler} \citep{2010ApJ...713L.176B,2010ApJ...723.1607H} and CoRoT \citep{2011A&A...525L...9M} as well as in the CoRoT target HD49385 \citep{2010A&A...515A..87D}. We expect to detect them in solar-like stars with longer time series and with higher SNR targets.

A deeper analysis of the characteristics of the p modes of these two stars, such as heights, linewidths, and amplitudes, will be presented in another paper (Handberg et al., in preparation).

With the global parameters of the p modes and the revised values of $T_{\rm eff}$, we can obtain a first estimation of the mean density,  radius and mass of these stars by using the scaling laws \citep{1995A&A...293...87K}, which are thus model independent. We estimate the mean density of these stars: $\langle \rho \rangle$~=~0.173~$\pm$~0.007~g/cm$^3$ for KIC11395018 and $\langle \rho \rangle$~=~0.132~$\pm$~0.008~g/cm$^3$  for KIC~11234888. We remind here the scaling laws:

\begin{equation}
\frac{R}{R_\odot}=\Big( \frac{135}{\langle \Delta \nu \rangle}\Big)^2 \Big(\frac{\nu_{\rm max}}{3050} \Big) \Big(\frac{T_{\rm eff}}{5777} \Big)^{1/2}\\
\end{equation}

\begin{equation}
\frac{M}{M_\odot}=\Big(\frac{135}{\langle \Delta \nu \rangle} \Big)^4 \Big(\frac{\nu_{\rm max}}{3050} \Big)^3 \Big(\frac{T_{\rm eff}}{5777} \Big)^{3/2}. \\
\end{equation}

For KIC~11395018, we obtain: M~=~1.25~$\pm$~0.24~$M_{\odot}$ and R=2.15~$\pm$~0.21~$R_{\odot}$, while for KIC~11234888, we have M~=~1.33~$\pm$~0.26~$M_{\odot}$ and R=2.4$\pm$~0.24~$R_{\odot}$. 

A detailed study based on stellar models, which is out of the scope of this work, is now possible. The modeling of this star requires the combination of global oscillation parameters, sets of frequencies, and atmospheric parameters (from spectroscopy), in order to retrieve very precise values of the stellar parameters, namely mass, radius, and age (Creevey et al. in preparation).

\acknowledgments
Funding for this Discovery mission is provided by NASAs Science Mission Directorate. The authors wish to thank the entire Kepler team, without whom these results would not be possible. We also thank all funding councils and agencies that have supported the activities of KASC Working Group 1, and the International Space Science Institute (ISSI). NCAR is supported by the National Science Foundation. SH also acknowledges financial support from the Netherlands Organisation for Scientific Research (NWO).
\bibliographystyle{apj} 
\bibliography{/Users/Savita/Documents/BIBLIO_sav} 

\clearpage

\begin{deluxetable}{lcccccc}
\tabletypesize{\scriptsize}

\tablecaption{Granulation and photon noise parameters of KIC~11395018 and KIC~11234888, fitted on the power spectrum density using the filtered Q01234 data set.\label{tbl-0}}
\tablewidth{0pt}
\tablehead{
&\colhead{$\tau_c$ [s]}& \colhead{$\sigma_c$ [ppm]} & \colhead{$\alpha_c$} & \colhead{$W$ [$\mathrm{ppm^2\,\mu Hz^{-1}}$]} & \colhead{$a$ ($\rm{ppm}^2 \mu$Hz$^{-1+b}$)} & \colhead{$b$}\\
}
\startdata
KIC~11395018 & $698\pm 33$ & $93.1\pm 0.9$ & $2.02\pm 0.06$ & $3.69 \pm 0.02$ & $1563.28 \pm 1142.16$ & $1.84 \pm 0.33$\\
KIC~11234888 & $869\pm 89$ & $88.3\pm 1.7$ & $1.68\pm 0.11$ & $12.75\pm 0.06$ &  $4130.78 \pm 5129.70$ & $1.67 \pm 0.56$ \\
\enddata
\end{deluxetable}

\clearpage

\begin{table*}
\begin{center}
\caption{Global parameters of KIC~11395018 and KIC~11234888 using Q01234 data.\label{tbl-1}}
\begin{tabular}{ccccccc}
\tableline\tableline
Star &  $\Delta \nu$ ($\mu$Hz)&  $\nu_{\rm max}$ ($\mu$Hz) &  $\langle \delta_{02} \rangle$ ($\mu$Hz) &$\langle \Gamma \rangle$ ($\mu$Hz) & P$_{\rm rot}$ (days)\\
\tableline
KIC~11395018 & 47.76 ~$\pm$~0.99& 830~$\pm$~48 & 4.12~$\pm$~0.03 & 0.84~$\pm$~0.02& 36\\
KIC~11234888 & 41.74~$\pm$~0.94& 675~$\pm$~42  & 2.38~$\pm$~0.19 & 0.86~$\pm$~0.06& 19-27\\

\tableline
\end{tabular}
\end{center}
\end{table*}

\begin{deluxetable}{ccccccc}
\tabletypesize{\scriptsize}

\tablecaption{Summary of the fitting methods.\label{tbl-6}}
\tablewidth{0pt}
\tablehead{
\colhead{Fitter ID} &  \colhead{ Method}  & \colhead{ Splittings} & \colhead{ Angle} & \colhead{ Heights} &  \colhead{ Linewidth} & \colhead{ Fit mixed mode}
}
\startdata
AAU & MCMC  Global & Free  & Free & H($l$=1)/H($l$=0)=1.48 & Free for $l$=0& No \\
 & & & & H($l$=2)/H($l$=0)=0.5& Interpolation for other modes  & \\
 & & & & & with $l$=0 mode\\
A2Z RG & MLE Global & Free & Free & H($l$=1)/H($l$=0)=1.5&Same for each order& Yes\\
& MAP& Guess 1~$\mu$Hz& Guess 45$^\circ$&  H($l$=2)/H($l$=0)=0.5 \\
A2Z CR & MLE Global& Fixed to 0~$\mu$Hz & Fixed to 0$^\circ$ & H($l$=1)/H($l$=0)=1.5 & Same for each order &No\\
& MAP& & &  H($l$=2)/H($l$=0)=0.5  & Guess 1~$\mu$Hz\\
A2Z DS & MLE  Local& Free & Free  &Free & Same for each order&No \\
 & &  Guess 0.8~$\mu$Hz & Guess 45$^\circ$\\
IAS OB & MCMC  Global & Free   & Free & Free & Free & No\\
IAS TA &  MLE Global & Fixed to 0~$\mu$Hz &- &  H($l$=1)/H($l$=0)=1.5 & Same for each order & Yes\\
& & & &  H($l$=2)/H($l$=0)=0.5 \\
IAS PG &  MLE Global & Free & Free &H($l$=1)/H($l$=0)=1.5 &Free  &No\\
 & MAP& Guess 1~$\mu$Hz& Guess 45$^\circ$&  H($l$=2)/H($l$=0)=0.5 & Guess 2~$\mu$Hz \\
OCT &  MLE Global & -& -& Free & Same for all modes & No\\
ORK & CLEAN  & -& -& -& -&No\\
QML & MLE  Global & Free & Free & Free&   One linewidth per overtone& No\\
SYD & Peaks in   & -&- & Free & -&Yes\\
 & smoothed spectra \\
 \enddata
\end{deluxetable}

\begin{table*}[htb]
\begin{center}
\caption{Minimal and maximal lists of frequencies for KIC~11395018 in $\mu$Hz obtained with eight months of data.\label{tbl-2}}
\begin{tabular}{cccc}
\tableline\tableline
  Order &$l$ = 0 & $l$ =1 & $l$ = 2  \\
\tableline
12 &... & 667.05 $\pm$0.22\tablenotemark{b} & 631.19 $\pm$ 1.36\tablenotemark{b} \\
13 & 686.66 $\pm$ 0.32 & 707.66 $\pm$0.19 & 680.88 $\pm$0.45\\
14 & 732.37 $\pm$ 0.18 & 763.99$\pm$0.18; 740.29$\pm$0.17\tablenotemark{a} & 727.78 $\pm$0.30\\
15& 779.54 $\pm$ 0.14 & 805.74$\pm$0.13 & 774.92 $\pm$0.16\\
16 & 827.55 $\pm$ 0.15 & 851.37 $\pm$0.11   & 823.50 $\pm$0.16\\
17 & 875.40 $\pm$ 0.16 & 897.50 $\pm$0.15 & 871.29 $\pm$0.21\\
18 & 923.16 $\pm$ 0.19 & 940.50 $\pm$0.15 & 918.10 $\pm$0.28\\
19 & 971.05 $\pm$ 0.28 & 997.91 $\pm$0.33 & 965.83 $\pm$0.23\\
20 & ... & ... & 1016.61 $\pm$0.73\tablenotemark{b,c}\\
\tableline
\end{tabular}
\tablenotetext{a} {Mixed mode}
\tablenotetext{b} {Part of the maximal list}
\tablenotetext{c} {Uncertain identification, could be an $l$=0}
\end{center}
\end{table*}

\clearpage

\begin{table*}[htb]
\begin{center}
\caption{Lists of frequencies for KIC~11234888 in $\mu$Hz obtained with eight months of data.\label{tbl-3}}
\begin{tabular}{cccc}
\tableline\tableline
  Order &$l$ = 0 & $l$ =1 & $l$ = 2  \\
\tableline
11 & ... & 506.72 $\pm$0.20& ... \\
12 & ... & 563.30 $\pm$0.14 & 582.84 $\pm$0.21\tablenotemark{a}\\
13 & 627.67 $\pm$ 0.19 & 594.83 $\pm$0.16 & 624.65 $\pm$0.18\\
14 & 669.35 $\pm$ 0.16 & ... & ...\\
15 & 711.63 $\pm$ 0.15 & 686.34 $\pm$0.17 & 708.67 $\pm$0.19\\
16 & 753.64 $\pm$ 0.20 & 741.10 $\pm$0.18 & 751.84 $\pm$0.21\\
17 & 794.56 $\pm$ 0.20 &  ... & ...\\
18 & 836.83 $\pm$ 0.21 & 815.43 $\pm$0.21 & ...\\
19 & 877.80 $\pm$ 0.22\tablenotemark{a} & 855.67 $\pm$0.21&... \\
\tableline
\end{tabular}
\tablenotetext{a} {Part of the maximal list}
\end{center}
\end{table*}

\end{document}